\documentclass[prd,amsfonts,amsmath,twocolumn,nofootinbib]{revtex4}

\usepackage{nicefrac}

\usepackage{color}
\usepackage{graphicx} 
\usepackage{epstopdf}%

\newcommand {\be} {\begin{equation}} 
\newcommand {\ba}{\begin{eqnarray}} 
\newcommand {\ee} {\end{equation}} 
\newcommand{\ea} {\end{eqnarray}}

\renewcommand{\epsilon}{\varepsilon}

\begin{document}

\title{Resonance Region Structure Functions and Parity Violating Deep Inelastic Scattering}

\author{Carl E. Carlson}

\author{Benjamin C. Rislow}

\affiliation{Department of Physics, College of William and Mary, Williamsburg, VA 23187, USA}

\date{\today}

\begin{abstract}
The primary motive of parity violating deep inelastic scattering experiments has been to test the standard model, particularly the axial couplings to the quarks,  in the scaling region.  The measurements can also test for the validity of models for the off-diagonal structure functions $F_{1,2,3}^{\gamma Z}(x,Q^2)$ in the resonance region.   The off-diagonal structure functions are important for the accurate calculation of the $\gamma Z$-box correction to the weak charge of the proton. Currently,  with no data to determine $F_{1,2,3}^{\gamma Z}(x,Q^2)$ directly,  models are constructed by modifying existing fits to electromagnetic data.   We present the asymmetry value for deuteron and proton target predicted by  several different $F_{1,2,3}^{\gamma Z}(x,Q^2)$ models, and demonstrate that there are notable disagreements.
\end{abstract}

\maketitle

%\baselineskip 13 pt

%%%%%%%%%%%%%%%%%%%%%%%%%%%%%%%%%%%%%%%%%%%%

\section{Introduction}			\label{sec:intro}

%%%%%%%%%%%%%%%%%%%%%%%%%%%%%%%%%%%%%%%%%%%%

%\nicefrac{3}{8}

Parity violating deep inelastic scattering experiments (PVDIS) in the past~\cite{Prescott:1978tm} and still today~\cite{PVDIS} have been motivated by the desire to search for physics beyond the Standard Model in lepton-quark neutral current interactions.  Consistent with the focus on interpreting the results in terms of quark couplings, the kinematics are chosen to mainly lie in the scaling region.

However, the recent interest in the larger than expected $\gamma$-$Z$ box correction to the Qweak experiment~\cite{Rajotte:2011nn} provides a further motivation for PVDIS, particularly for any data that may lie in the resonance region.  
QWeak is also an experiment designed to test the standard model, using elastic electron-proton scattering with polarized electrons, and obtaining an accurate result requires good knowledge of higher order corrections.  Hence the interest in the $\gamma$-$Z$ boxes.   

Both PVDIS and the $\gamma$-$Z$ box calculations are dependent upon the off-diagonal structure functions $F_{1,2,3}^{\gamma Z}$, defined from the spin-averaged tensor
\begin{align}
\label{Wmunu}
W_{\mu \nu}^{\gamma Z}&=\int \frac{d^4 \xi}{4\pi} \, e^{iq\xi}\langle ps\left|J_{Z\mu}(\xi)J_{\gamma\nu}(0)+J_{\gamma\mu}(\xi)J_{Z\nu}(0\right|ps\rangle
\nonumber \\
&= \left(-g_{\mu \nu}+\frac{q_\mu q_\nu}{q^2} \right) F^{\gamma Z}_1(x,Q^2)+\frac{p_\mu p_\nu}{p\cdot q}F^{\gamma Z}_2(x,Q^2)
       \nonumber\\
& -i\epsilon_{\mu\nu\alpha\beta}\frac{q^\alpha p^\beta}{2p\cdot q}F^{\gamma Z}_3(x,Q^2)    .
\end{align}
Functions $F^{\gamma Z}_{1,2}$ come from the vector terms in the $Z$-boson current and $F^{\gamma Z}_3$ comes from the axial vector terms;  we may call them $Z_V$ and $Z_A$ couplings for brevity.

\begin{figure}[b]
\begin{center}
\includegraphics[width = 3.37 in]{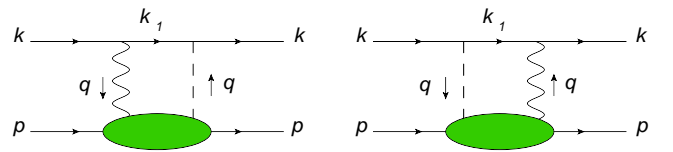}
\caption{The $\gamma$-$Z$ box diagrams.}
\label{fig:mgz}
\end{center}
\end{figure}

The $\gamma$-$Z$ box, Fig.~\ref{fig:mgz},  can be calculated dispersively in terms of the structure functions $F_{1,2,3}^{\gamma Z}$~\cite{Gorchtein:2008px,Sibirtsev:2010zg,Rislow:2010vi,Gorchtein:2011mz,Blunden:2011rd}.  The surprising recent result~\cite{Gorchtein:2008px} was that the vector $Z$-boson contributions were larger than expected and numerically comparable to the axial contributions, which had comprised the bulk of the earlier estimates (see for example~\cite{Erler:2003yk}).   In the scaling region and with the standard model one obtains the $\gamma$$Z$ structure functions from parton distribution functions. However, the contributions from 
$F_{1,2}^{\gamma Z}$ are given as integrals that have their main support in the resonance region and at moderate $Q^2$.  In the absence of data for these structure functions, one has obtained them using model-based modification of data from other channels,  in particular starting from the standard electromagnetic structure functions, here called 
$F_{1,2}^{\gamma\gamma}$.    Regarding the $F_{3}^{\gamma Z}$ integrals, the main support is at higher energy and higher $Q^2$, so that known parton distribution functions can be used to obtain the bulk of the axial contributions.  However, one still wants to know $F_3^{\gamma Z}$ in the resonance region.   In principle, this can be obtained from the charge current reaction data, since only the Weak axial current is involved.  However, Weak interaction resonance region data is scarce so that modeling is still needed. 

%%%%%%%%%%%%%

%%%%%%%%%%%%%

PVDIS allows a direct measurement of the $\gamma$$Z$ structure functions.  The PVDIS asymmetry is given by~\cite{Nakamura:2010zzi}
\begin{align}
&A_{PVDIS} = \frac{ \sigma^{NC}(\lambda=1/2) -  \sigma^{NC}(\lambda=-1/2) }
		{ \sigma^{NC}(\lambda=1/2) +  \sigma^{NC}(\lambda=-1/2) }	\nonumber\\
	& = g_A^e \frac{G_F Q^2}{ 2\sqrt{2}\pi\alpha}				
													\nonumber\\
&\times	\frac{  xy^2 F_1^{\gamma Z} 
		+  \left( 1 - y - \frac{x^2 y^2 M^2}{Q^2} \right) F_2^{\gamma Z} 
		+ \frac{g_V^e}{g_A^e} \left( y - \frac{y^2}{2} \right) xF_3^{\gamma Z} }
		{ x y^2 F_1^{\gamma\gamma}  
		+ \left( 1 - y - \frac{x^2 y^2 M^2}{Q^2} \right)  F_2^{\gamma\gamma}}			\nonumber\\[1.2ex]
	&= g_A^e \frac{G_F Q^2}{ 2\sqrt{2}\pi\alpha}
													\nonumber\\
&\times	\frac{ 2  \sin^2 \frac{\theta}{2} F_1^{\gamma Z}
		+  \frac{M}{\nu} \cos^2 \frac{\theta}{2} F_2^{\gamma Z}
+ \frac{g_V^e}{g_A^e} \frac{2(E+E')}{\nu} \sin^2\frac{\theta}{2} \, F_3^{\gamma Z} }
		{ 2 \sin^2 \frac{\theta}{2} F_1^{\gamma\gamma} 
		+ \frac{M}{\nu} \cos^2 \frac{\theta}{2}  F_2^{\gamma\gamma} }	\,.
\end{align}
Here, $\lambda$ is the incoming lepton helicity, $E$ and $E'$ are the incoming and outgoing lepton energy in the target rest frame, $\nu=E-E'$, $\theta$ is the lepton scattering angle, $x=Q^2/(2M\nu)$, $y=\nu/E$, and in the standard model, $g_A^e = -1/2$ and $g_V^e = -1/2+ 2 \sin^2\theta_W$, where $\theta_W$ is the Weinberg angle.

In principle, measuring the asymmetry over a range of angles, initial energies, and $Q^2$ allows a full determination of the structure functions.  In practice, at least at the outset, the data will be limited, and so it will be useful to predict the asymmetries in the resonance region using the models developed for evaluating the $\gamma$$Z$ boxes, and then use the data to spot-check the models.  It is worth mentioning at the outset that for most relevant kinematics the numerical contribution of the $F_3^{\gamma Z}$ term will be small.

Although this article is focused on the resonance region,  for the purpose of contrast and comment we give the PVDIS asymmetry formula specialized to the scaling region,
\begin{widetext}
\begin{equation}
A_{PVDIS} = \frac{3 G_F Q^2}{ 2\sqrt{2}\pi\alpha}
\frac{2 C_{1u} (u_A+\bar u_A) - C_{1d} (d_A+\bar d_A + s_A+\bar s_A)
+ Y \left( 2C_{2u} (u_{A}- \bar u_A)  - C_{2d} (d_{A}-\bar d_A)  \right)}
{4 (u_A+\bar u_A) + d_A+\bar d_A + s_A+\bar s_A }		\,,
\end{equation}
\end{widetext}
where $q_A$ is the distribution function for quark $q$ in target $A$.  The $F_3^{\gamma Z}$ term has become the term containing $Y(y)$ and $C_{2q}$, where 
\begin{equation}
Y(y) = \frac{ 1 - (1-y)^2 }{ 1 + (1-y)^2 }	\,,
\end{equation}
and
\begin{equation}
C_{1q} = 2 g_A^e g_V^q  \quad , \quad
C_{2q} = 2 g_V^e g_A^q 	\,.
\end{equation}
In the standard model, $g_V^u = \frac{1}{2} - \frac{4}{3} \sin^2\theta_W$, $g_V^d = -\frac{1}{2} + \frac{2}{3} \sin^2\theta_W$, and $g_A^u = \frac{1}{2} = - g_A^d$.

The $F_3^{\gamma Z}$ contribution is largest for $x\to 1$ and $y\to 1$, where one has $Y \to 1$, and expects the antiquark and strange quark distributions to be very small.  For the standard model in this limit, one expects the $C_{2q}$ terms to be about $12$\% of the $C_{1q}$ terms in the deuteron, where one can let $u_A = d_A$.  For the proton, the effect is somewhat larger but dependent on the down to up quark ratio in the valence region ($x\to 1$).   Beyond the standard model searches will work in the scaling region and seek deviations from this result.

We, on the other hand, will here accept the standard model, and work in the resonance region and hope to learn about $F_{1,2}^{\gamma Z}$.

We continue by showing predictions for the PVDIS asymmetry in the resonance region based on several models that have been proposed and used in the $\gamma$$Z$ box calculations, and discussing the reasons for the differences among these model predictions, and then offer some conclusions.

%%%%%%%%%%%%%%%%%%%%%%%%%%%%%%%%%%%%%%%%%%%%

\section{Asymmetries in the resonance region}			\label{sec:res}

%%%%%%%%%%%%%%%%%%%%%%%%%%%%%%%%%%%%%%%%%%%%

\begin{figure*}[htbp]

\includegraphics[width = 70 mm]{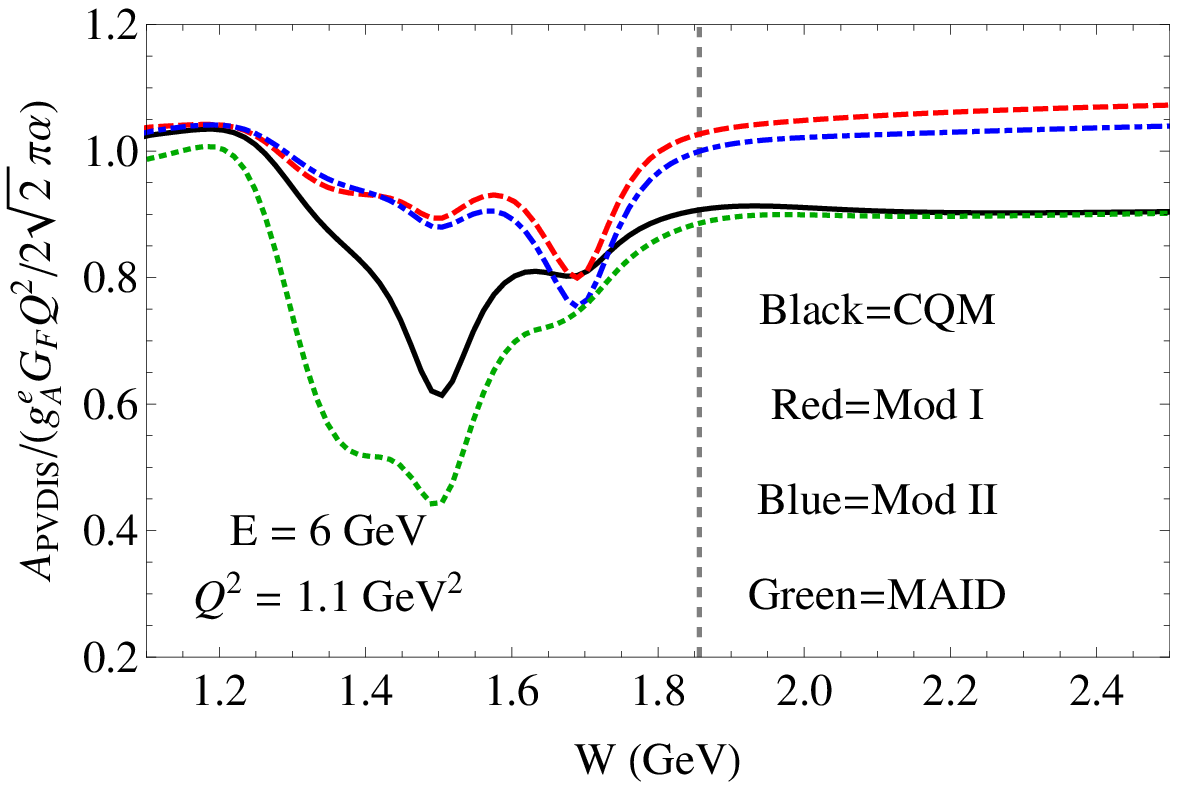} \hfil 
\includegraphics[width = 70 mm]{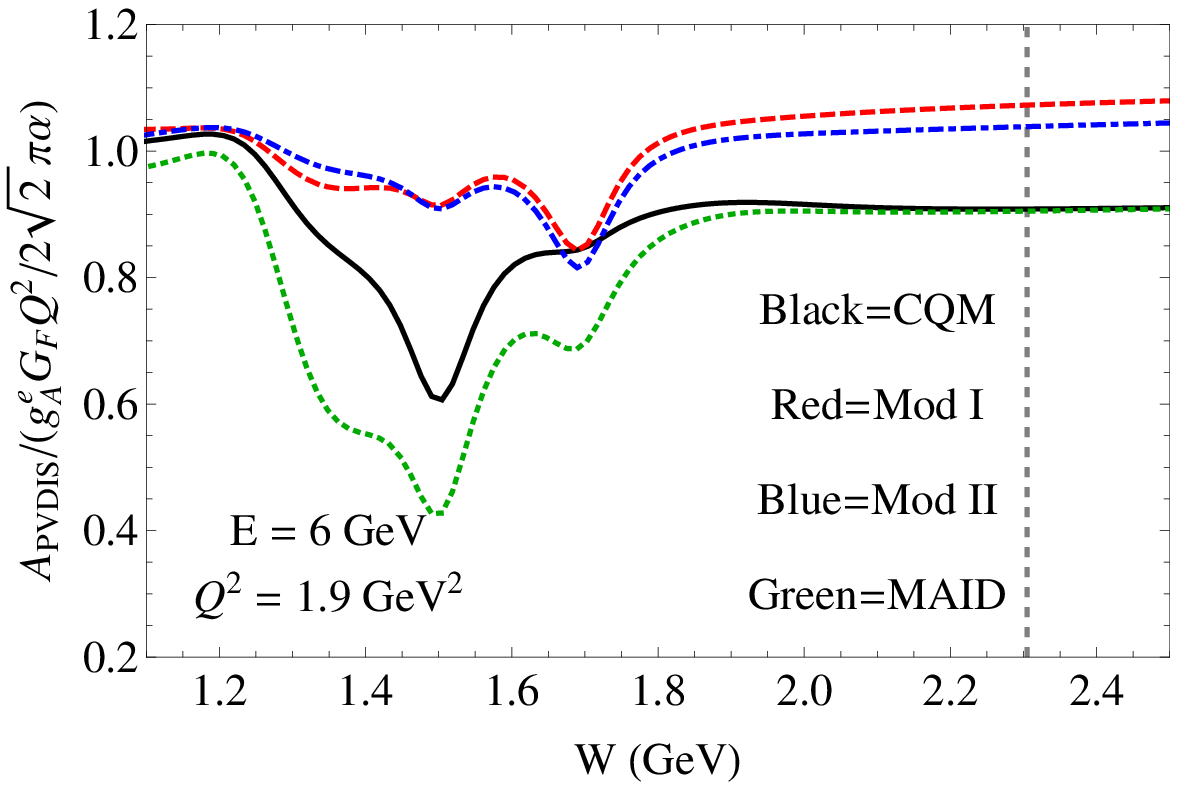}

\includegraphics[width = 70 mm]{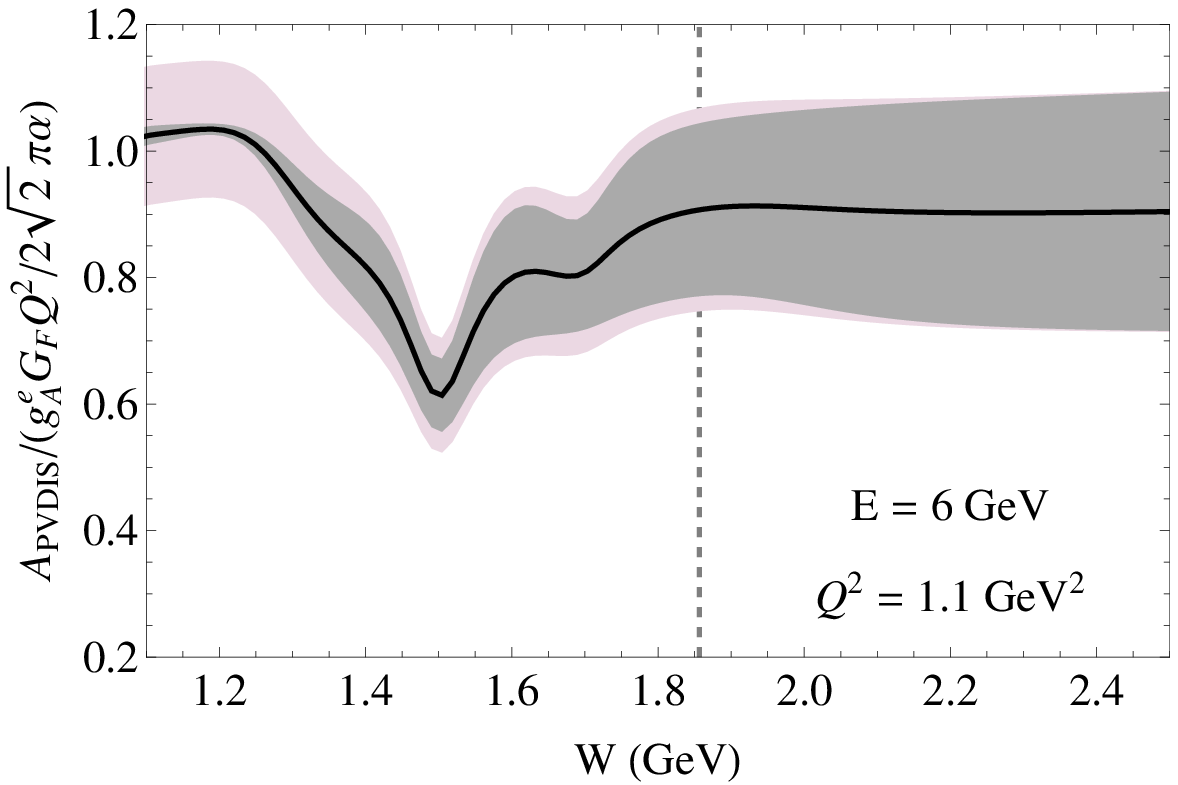} \hfil
\includegraphics[width = 70 mm]{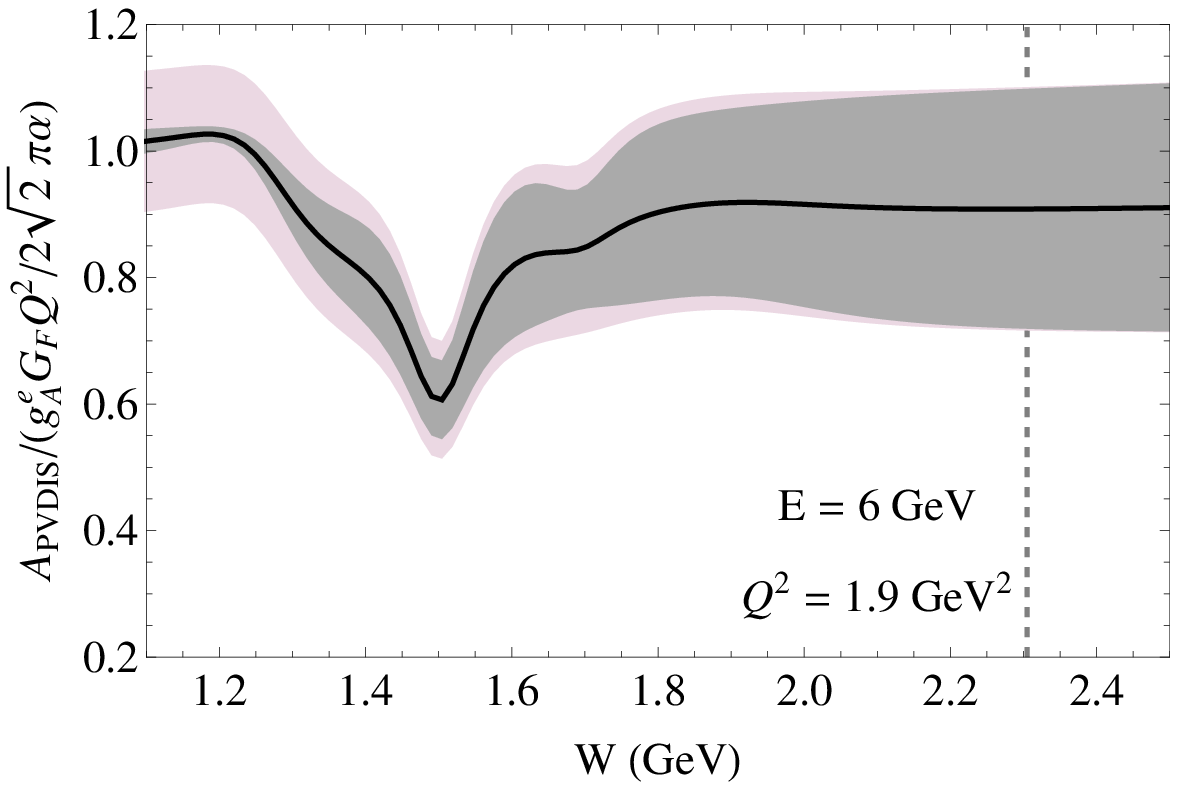}

\includegraphics[width = 70 mm]{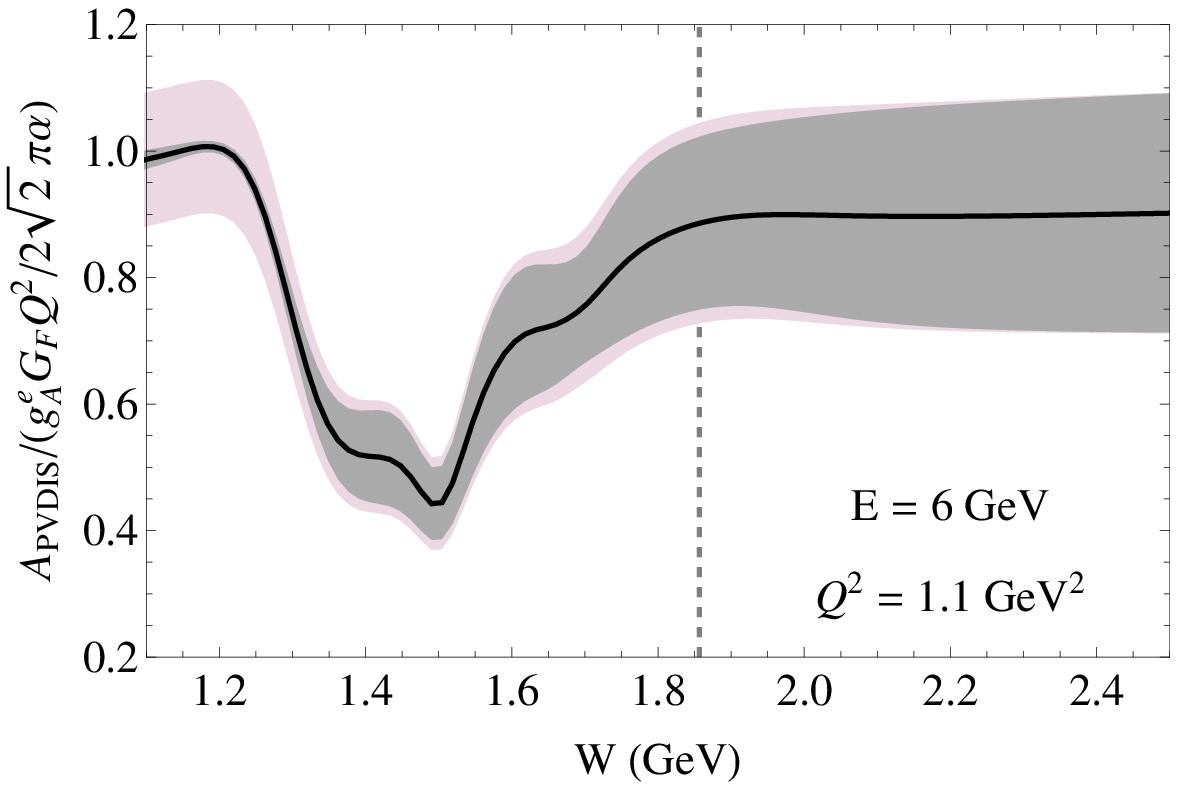} \hfil
\includegraphics[width = 70 mm]{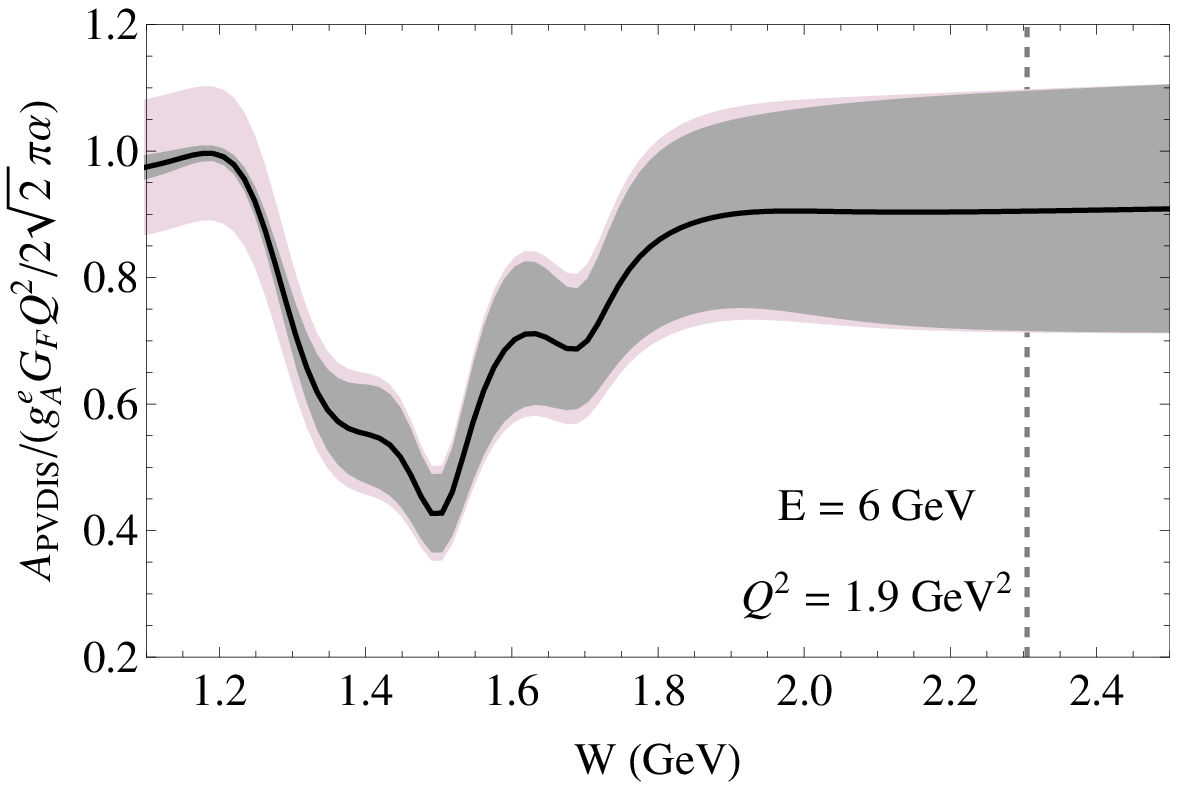}

\caption{Normalized proton asymmetry for $Q^2=1.1$ GeV$^2$ and $1.9$ GeV$^2$ as a function of $W$.   The top panels show results for several parameterizations,  one based on a constituent quark model based modification of the Christy-Bosted electromagnetic fits (black solid), one obtained using MAID fits to the resonance helicity amplitudes (green dotted), and two suggestions following Gorchtein \textit{et al.} (red dashed and blue dash-dot).  The middle and bottom panels give uncertainty limits for the constituent quark model and MAID  cases, respectively, with the grey band being uncertainty for the nonresonant terms alone, and the pink band adding uncertainty in the  resonant contributions to obtain the total.   The dashed vertical lines indicate the kinematic points for the 6 GeV PVDIS (deuteron) experiment;  each corresponds to $x \approx 0.3$.}

\label{fig:protonPVDIS}

\end{figure*}

\begin{figure*}[htbp]

\includegraphics[width = 70 mm]{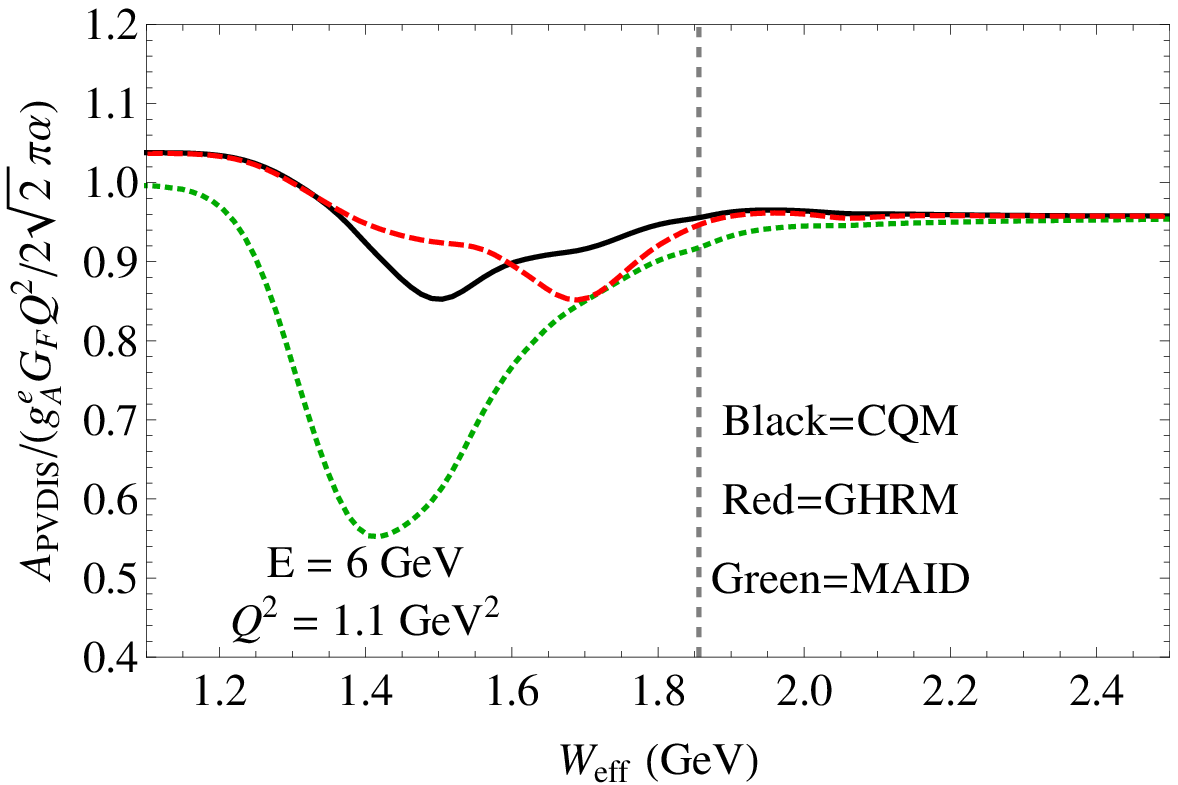} \hfil 
\includegraphics[width = 70 mm]{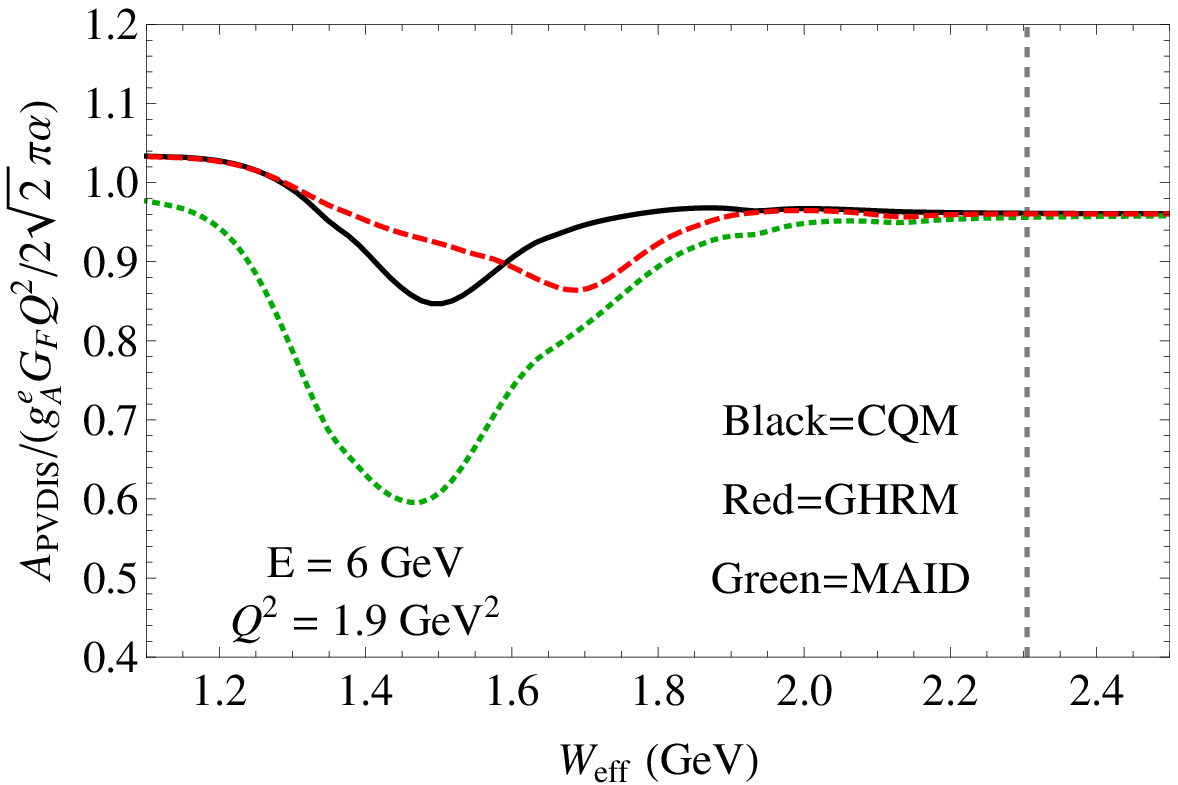}

\includegraphics[width = 70 mm]{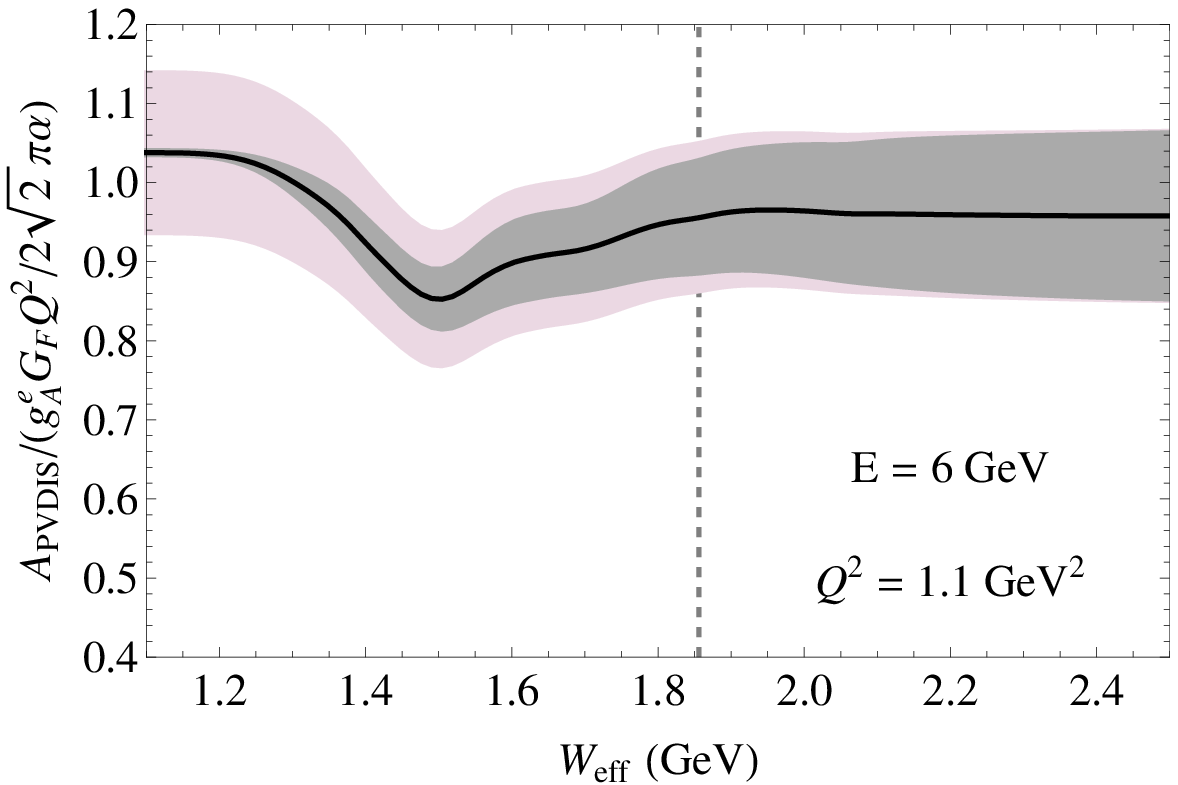} \hfil
\includegraphics[width = 70 mm]{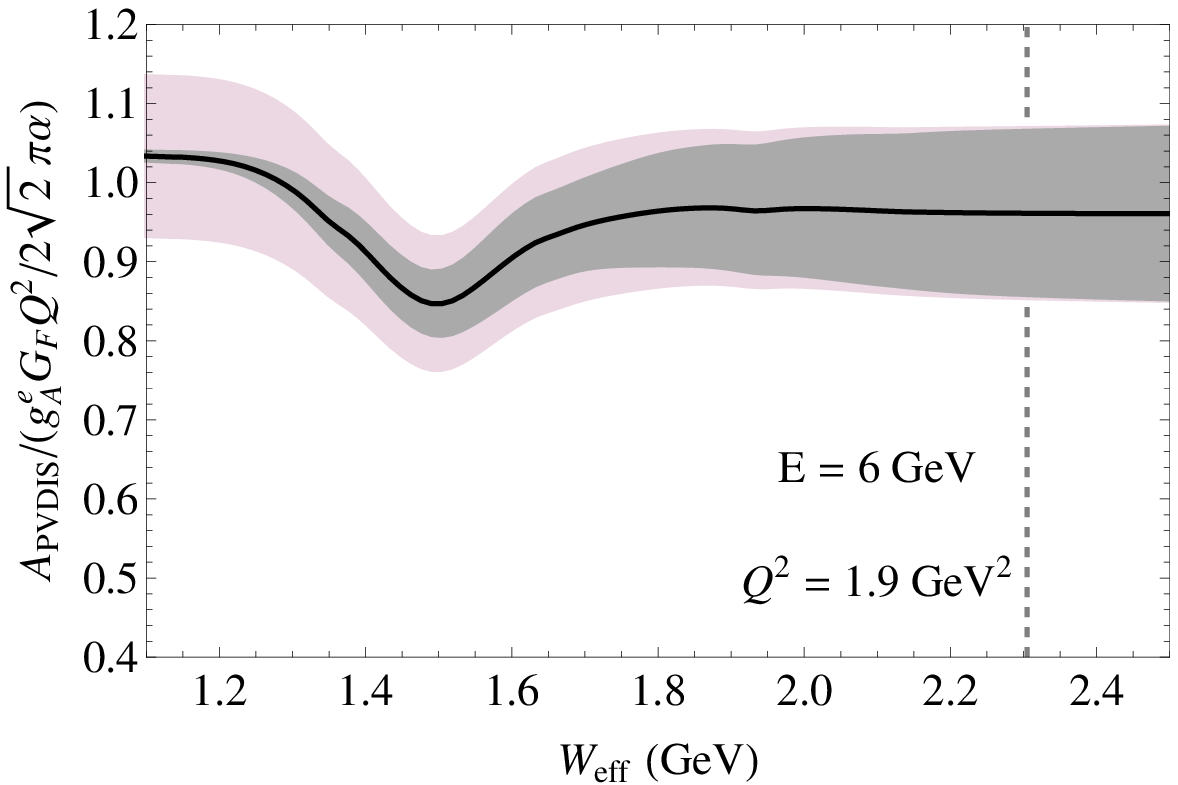}

\includegraphics[width = 70 mm]{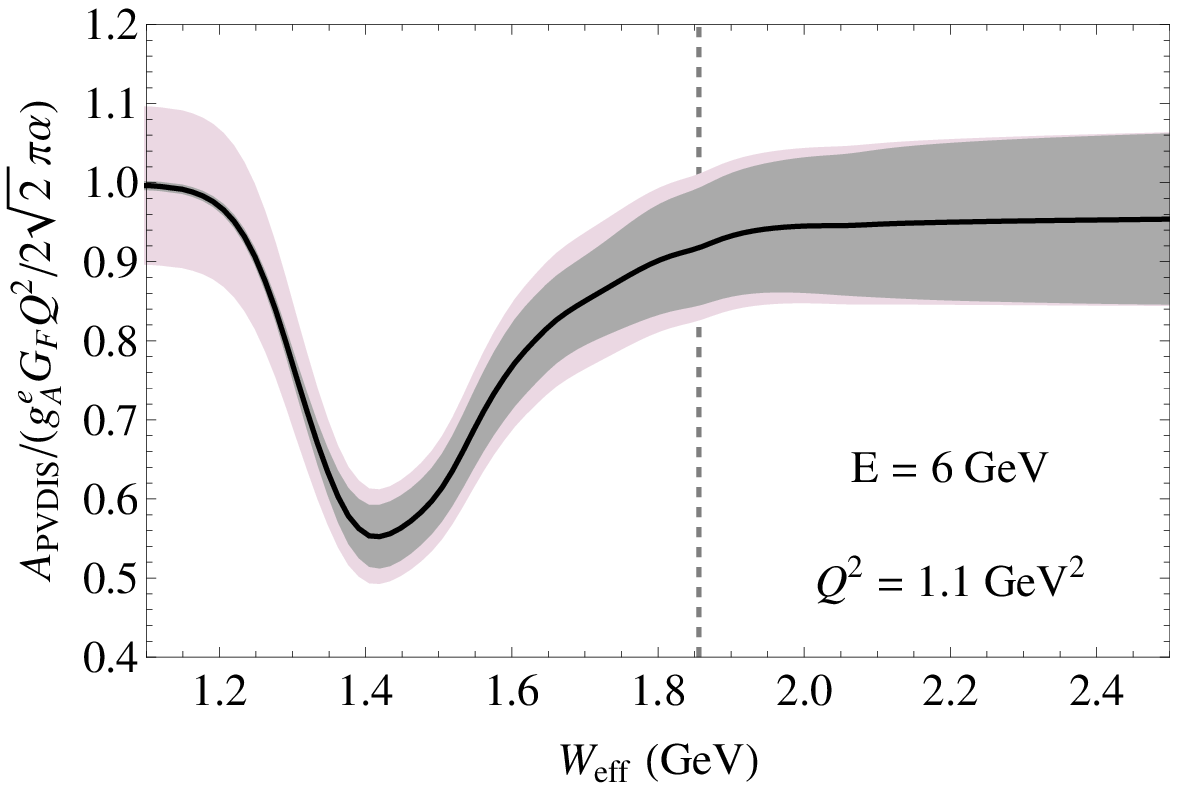} \hfil
\includegraphics[width = 70 mm]{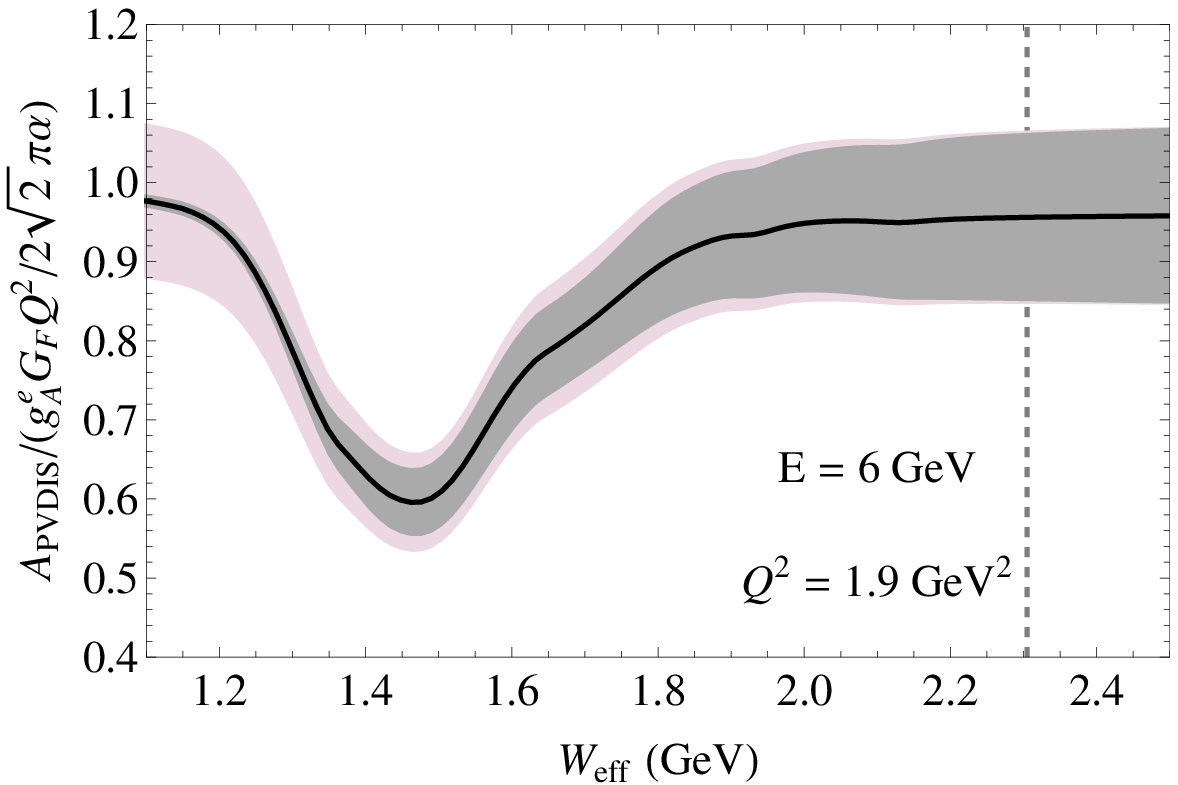}

\caption{Deuteron asymmetry for $Q^2=1.1$ GeV$^2$ and $1.9$ GeV$^2$ as a function of $W_{\rm eff}$, where $W_{\rm eff}^2 = M^2 +2M\nu - Q^2$.   The top panels show results for several parameterizations,  one based on a constituent quark model based modification of the Christy-Bosted electromagnetic fits (black solid), one obtained using MAID fits to the resonance helicity amplitudes (green dotted), and one using photoproduction data in the style of Gorchtein \textit{et al.} (red dashed).  The middle and bottom panels give uncertainty limits for the constituent quark model and MAID  cases, respectively, with the grey band being uncertainty for the nonresonant terms alone, and the pink band adding uncertainty in the  resonant contributions to obtain the total.   The dashed vertical lines again indicate the kinematic points for the 6 GeV PVDIS experiment;  each corresponds to $x \approx 0.3$.}
\label{fig:deuteronPVDIS}

\end{figure*}

Figures~\ref{fig:protonPVDIS} and~\ref{fig:deuteronPVDIS} show several examples of what models predict for the PVDIS asymmetry in the resonance region,  choosing for definiteness the two $Q^2$ values where the 6 GeV PVDIS experiment has been run.

The models for the $\gamma$$Z$ structure functions in the resonance region have been mainly discussed in the context of a proton target,   and so we begin with the proton in Fig.~\ref{fig:protonPVDIS}.   The top two panels show results for the PVDIS asymmetry that follow from four different models of how to convert the electromagnetic structure functions to the $\gamma$-$Z$ ones.  The vertical dashed line in each figure shows the value of $W$ which is targeted in the current (deuteron)  experiment.

%%%%%%%%%%%%%%%%%%%%%%
%%%%%%%%%%%%%%%%%%%

%%%%%%%%%%%%%%%%%%%
%%%%%%%%%%%%%%%%%%%%%%

Each of the $\gamma$$Z$ structure function models that we show have their roots in the Christy-Bosted fit~\cite{Christy:2007ve} to $\gamma$$\gamma$ structure function data.  To convert to the $\gamma$-$Z$ structure functions, each of the fit's seven electromagnetic resonance contributions is modified by a corrective ratio.  Explicitly, for the transverse contributions, one forms the ratios
\begin{equation}
\label{eq:int}
C_R = \frac{ 2 \sum A_\lambda(\gamma p \to R) A_\lambda(Z_V p \to R) }
	{ \sum | A_\lambda(\gamma p \to R) |^2 }
\end{equation}
(where $A_\lambda$ is the transverse helicity amplitude and $\lambda$ is the helicity magnitude of the resonance), and multiplies each respective electroproduction resonance contribution by its $C_R(Q^2)$.  The $\gamma$$Z$ structure function models differ in the way they obtain the corrective ratios, as well as in their handling of the nonresonant background.

In the top panels, the curve labeled ``CQM" follows from the model used by us in~\cite{Rislow:2010vi}.  The model is a constituent quark model for resonance production and is used to calculate helicity amplitudes for generic vector couplings.  Helicity amplitudes for $Z$-boson (photon) exchange are obtained by inserting the $g_V^q$ vector Weak charges ($e_q$ quark charges).  Many transitions receive contributions from both electric and magnetic excitations.  With extra information about the helicity behavior of the spin-3/2 resonances, one can determine the individual contributions from the electromagnetic data and then convert to the $Z_V$ amplitudes.  The background we use comes from the background part of the Christy-Bosted fits to inelastic electron scattering data,  modified for the $\gamma$$Z$ case.

The curves labeled ``Model I'' and ``Model II'' both follow from the models used in~\cite{Gorchtein:2011mz} (GHRM).  The analysis of the corrective ratio, Eq.~(\ref{eq:int}), for GHRM begins by considering the relation
\begin{align}			\label{eq:res}
2 \langle R^+ | J_\mu^{Z_V} | p \rangle = 
&	(1-4\sin^2\theta_W ) \langle R^+ | J_\mu^\gamma | p \rangle	\nonumber\\
&	- \langle R^0 | J_\mu^\gamma | n \rangle
	- \langle R^+ | \bar s \gamma_\mu s | p \rangle		\,.
\end{align}
The strange quark contribution is generally neglected, as suggested by data~\cite{Baunack:2009gy,:2009zu,:2011vp}, so that the vector $Z$-current helicity amplitudes can be gotten from neutron and proton data.  GHRM implement this only at the photoproduction point, using proton and neutron helicity amplitudes given by the Particle Data Group~\cite{Nakamura:2010zzi}.  They assume that the $Q^2$ dependence of the $Z_V$ and electromagnetic matrix elements are the same, so that they have $Q^2$ independent $C_R$'s.  They obtain resonant electromagnetic matrix elements from the resonant part of the Christy-Bosted fit~\cite{Christy:2007ve}, slightly modified to better match their background choices.  

The GHRM backgrounds come from two fits to electroproduction in a higher energy diffractive region, pushed to lower energy and isospin modified for the $\gamma$$Z$ case.  Model I is based on a color dipole model based fit of Cvetic \textit{et al.}~\cite{Cvetic:2001ie}.  In this model, the photon fluctuates into a $q$-$\bar q$ pair which interacts with the proton via gluon exchange.  Cvetic \textit{et al.} obtain a functional form with parameters, which are constrained by data, and a good fit is obtained for low $Q^2$, high energy structure functions.  GHRM~\cite{Gorchtein:2011mz} extrapolate this to model the background in the resonance region, and obtain the $\gamma$-$Z$ structure functions by changing the averaged photon-quark coupling to $Z$-quark coupling.  Model II is based on a generalized vector meson model based fit of Alwall and Ingelman~\cite{Alwall:2004wk}.  This model fits the electromagnetic structure functions at low $Q^2$ and high energy by relating them to the total photon proton cross sections and coupling the photon through vector meson intermediaries, given in terms of $\rho$, $\omega$, and $\phi$ plus a background or continuum contribution.  Similar to their Model I modifications, GHRM extrapolate the Model II fit to lower energies and use it as a background in the resonance region, first showing adequate fits to purely electromagnetic data and then modifying the fit for the $\gamma$-$Z$ case by considering $Z$-vector meson as well as $\gamma$-vector meson mixings.

The curve labeled ``MAID'' used the MAID fits to the resonance electroproduction helicity amplitudes with proton and neutron targets~\cite{Tiator:2011pw} to obtain the vector $Z$-boson matrix elements from Eq.~(\ref{eq:res}) and thence a fully $Q^2$ dependent corrective ratio.  These are then used to transform each of the seven resonant contributions in the Christy-Bosted fit into the $\gamma$$Z$ structure function terms.  The background is obtained as previously described when using the CQM model.  

The $F_3^{\gamma Z}$ terms have small numerical impact here,  at most 3\% of the total.  The $F_3^{\gamma Z}$ contributions are not big even in the scaling region under optimal kinematics, and at kinematics relevant to this paper, the function $Y(y)$ is well below unity and the presence of sea quark and strange quark contributions further reduce the relative size of the $F_3^{\gamma Z}$ terms.  To justify the numerical result, we need some estimate for $F_3^{\gamma Z}$.  For the resonance contribution we use a corrective ratio with a helicity amplitude for the $Z$-boson axial coupling and once again evaluate the amplitudes with our constituent quark model.  We have also obtained resonance contributions from the four-resonance fits of~\cite{Lalakulich:2006sw}, and find the difference it makes in our plots is very slight. For the background one takes guidance from the scaling region.  For very low-$x$, all light quarks and antiquarks have similar distributions, so that $F_3^{\gamma Z}$, which depends on differences $q(x) - \bar q(x)$, is about zero compared to $F_1^{\gamma\gamma}$.  Alternatively, in a valence quark dominated region where the up quark distribution is twice the down quark distribution, one has $F_3^{\gamma Z}/F_1^{\gamma\gamma} = 10/3$.  We take the average of the two limits, taking $F_3^{\gamma Z}$ for the proton to be 5/3 the Christy-Bosted $F_1^{\gamma\gamma}$ (with 100\% uncertainty bounds on this term, when we discuss uncertainty bounds below, to accommodate the two limits).

One can see the results of the various models differ.  This has much to do with the $Q^2$ dependences of the resonance amplitudes in the different models.  Note the deep dip in the quark model in the second resonance region contrasting with the deep dip in the third resonance region for the GHRM predicted asymmetry.  GHRM sets the ratio of the vector $Z$ to electromagnetic matrix element by their values at $Q^2=0$.  The quark model considers that both the second and third resonance regions have a spin-3/2 or higher resonance, which each have two helicity amplitudes and these amplitudes have different $Q^2$ falloffs. The weightings of the two amplitudes changes when switching from the photon to the $Z$-boson case, which leads to a different overall falloff with $Q^2$ for the $Z$-boson contributions, faster for the second resonance region and slower for the third resonance region.

The deep dip seen in the MAID based asymmetry is due to the behavior of the Roper resonance in the MAID model.  In the MAID fits, the proton transverse helicity amplitude changes sign, at about $\nicefrac{2}{3}$ GeV$^2$, while the neutron does not, leading to a sign change in some of the interference terms giving the $\gamma$$Z$ structure functions following from Eq.~(\ref{eq:int}).  The negative contribution from the Roper leads to the dip.   The GHRM and the quark model (as it happens for the Roper) have a $Q^2$ independent electromagnetic to $Z$ boson amplitude and hence no sign change and no dip in the Roper region.

For the uncertainty limits of the smooth non-resonant background to $F_{1,2}^{\gamma Z}$ in the quark model, we take guidance from scattering off collections of quarks with scant final state interactions.    In a full $SU_f(3)$ limit, where all light quarks are equally likely and which may be pertinent in a high-energy $x = Q^2/(2M\nu)\to 0$ limit, one has $F_{1,2}^{\gamma Z}/F_{1,2}^{\gamma \gamma} = 1 + Q_W^{p,LO}$.  In a valence quark limit with SU(6) wave functions, one gets $(2/3 + Q_W^{p,LO})$ for the same ratio.  The latter is better at high-$x$ and the former is better at low-$x$ and we take the mean for our central curve, and use the extremes to set the uncertainty estimates which are shown in the four lower panels of Fig~\ref{fig:protonPVDIS}.  Regarding the conversion of the resonances from electromagnetic to $Z$-boson matrix elements,  we assigned a 10\% uncertainty in each matrix element~\cite{Rislow:2010vi}.  The uncertainties are shown in the middle panels of Fig.~\ref{fig:protonPVDIS} are for the CQM,  with the grey band showing the uncertainty limits for the nonresonant terms, and the pink band adding uncertainty from the resonant contributions to obtain the total.  The bottom panels show similar bands for the MAID based asymmetry predictions;  results for GHRM are qualitatively similar.

For the deuteron asymmetry, Fig.~\ref{fig:deuteronPVDIS}, we modify the Bosted-Christy deuteron fits to electromagnetic structure function data~\cite{Bosted:2007xd}.  Extending the quark model to obtain the $Z$ current matrix elements on a neutron target is of course straightforward.  Also, the Weak isospin rotation giving the $Z$ boson on neutron matrix elements reflects the proton case,
\begin{align}			\label{eq:resn}
2 \langle R^0 | J_\mu^{Z_V} | n \rangle = 
&	(1-4\sin^2\theta_W ) \langle R^0 | J_\mu^\gamma | n \rangle	\nonumber\\
&	- \langle R^+ | J_\mu^\gamma | p \rangle
	- \langle R^0 | \bar s \gamma_\mu s | n \rangle	\,.
\end{align}

The constituent quark model and MAID based treatments of the resonances are thus both straightforward to extend to the deuteron.  However, a problem in extending the GHRM treatment to the deuteron is that we have no equivalent of the color dipole model~\cite{Cvetic:2001ie} or generalized vector dominance~\cite{Alwall:2004wk} fits, which were used for the background, to the deuteron.  However, we can, as we do for the quark model and MAID, use the Bosted-Christy deuteron background, suitably modified for the $\gamma$$Z$ case, and combine with the GHRM resonance fits to do a fit in their style.   

The $F_3^{\gamma Z}$ structure function is again a small contribution, and we obtain an estimate of it in the same way as we did for the proton target.  The background part of $F_3^{\gamma Z}/F_1^{\gamma\gamma}$ is still essentially zero in high-energy low-$x$ limit, but becomes $18/5$ for the deuteron in the valence only limit.  We again take the average of the two limits, taking $F_3^{\gamma Z}$ for the proton to be 9/5 the Bosted-Christy $F_1^{\gamma\gamma}$ for the deuteron, with 100\% uncertainty bounds to accommodate the two limits.

Comments about the reasons for the differences between the quark, GHRM, and MAID based treatments are similar to those made for the proton case.

%%%%%%%%%%%%%%%%%%%%%%%%%%%%%%%%%%%%%%%%%%%%

\section{Conclusion}			\label{sec:end}

%%%%%%%%%%%%%%%%%%%%%%%%%%%%%%%%%%%%%%%%%%%%

Parity violating deep inelastic asymmetry measurements were proposed as a source of interesting and useful information about the $\gamma$$Z$ interference structure functions in the scaling region.  The intent was to use a deuteron target to minimize the uncertainty in removing effects of the large $F_1^{\gamma Z}$ and $F_2^{\gamma Z}$ terms and hence isolate $F_3^{\gamma Z}$.  In the latter, the $Z$-quark coupling is axial vector, and the goal was to measure if the coupling was in accord with the standard model.

On the other hand, if the standard model is valid, PVDIS provides a way to measure the $\gamma$$Z$ structure functions in any kinematic region.   In particular, the results in the resonance region would be very interesting and useful.   The interest in this region is to measure or constrain the $F_{1,2}^{\gamma Z}$ structure functions, and the smallness of the $F_3^{\gamma Z}$ becomes an advantage.  The resonance region $F_{1,2}^{\gamma Z}$ are not in fact well predicted from existing data in other channels.  Several models are available, and we have shown the PVDIS asymmetries that follow from several of them, and note that the predicted asymmetries are fairly distinct, differing by several tens of percent in some kinematic regions.  

The $\gamma$$Z$ structure functions have an immediate application, which is in the calculation of the $\gamma$$Z$ box corrections to elastic parity violating electron-proton scattering~\cite{Gorchtein:2008px,Sibirtsev:2010zg,Rislow:2010vi,Gorchtein:2011mz}.  The corrections are calculated using dispersion relations, and the numerical contributions from the $Z$ boson having vector or axial interactions is about the same.  However,  the weightings within the integrals are such that when the $Z$ has vector couplings, which give the $F_{1,2}^{\gamma Z}$,  the main support comes from the resonance region.  Hence the direct interest in measuring these functions there.   For the $Z$-boson axial couplings, which give $F_3^{\gamma Z}$, most of the $\gamma$$Z$ box contribution comes from the scaling region~\cite{Blunden:2011rd}.  Hence there is less need from the present viewpoint for measuring $F_3^{\gamma Z}$ in the resonance region, and the smallness of the $F_3^{\gamma Z}$ contribution to PVDIS can even be counted as an advantage.

%%%%%%%%%%%%%%%%%%%%%%%%%%%%%%%

\begin{acknowledgments}

We thank the National Science Foundation for support under Grant PHY-0855618 and thank Wally Melnitchouk and Kent Pashke for helpful comments.

\end{acknowledgments}

\bibliography{PVDIS}

\end{document}